\def\1{\'{\i}}
\def\beq{\begin{equation}}
\def\eeq{\end{equation}}
\def\bea{\begin{eqnarray}}
\def\eea{\end{eqnarray}}
\def\bed{\begin{displaymath}}
\def\eed{\end{displaymath}}

\documentstyle[preprint,pre,aps,epsf]{revtex}
\begin{document}
\preprint{}

\title{\bf Unstable dimension variability and synchronization of chaotic systems}

\author{Ricardo L. Viana $^1$ and Celso Grebogi $^2$}
\address{1. Departamento de F\'{\i}sica, Universidade Federal do Paran\'a, 
	81531-990, Curitiba, PR, Brazil. \\
	2. Department of Mathematics, Institute for Plasma Research and 
	Institute for Physical Science \\
	and Technology, University of Maryland, College Park, Maryland 20742}

\date{\today}	
\maketitle

\begin{abstract}
An aspect of the synchronization dynamics is investigated in this work. We argue analytically and confirm numerically that the chaotic dynamics on the synchronization manifold exhibits unstable dimension variability. Unstable dimension variability is a cause of severe modeling difficulty for physical phenomena, since trajectories obtained from the mathematical model may not be related to trajectories of the actual system. We present an example of unstable dimension variability occurring in a system of two coupled chaotic maps, considering the dynamics in the synchronization manifold and its corresponding transversal direction, where a tongue-like structure is formed. The unstable dimension variability is revealed in the statistical distribution of the finite-time transversal Lyapunov exponent, having both negative and positive values.

\noindent05.45.Xt,05.45.Pq
\end{abstract}

\narrowtext

\section{Introduction}

We usually classify dynamical systems into deterministic and stochastic ones. Deterministic systems are characterized by a set of prescribed mathematical rules which evolve the dynamical variables in time. For these kinds of models we can numerically generate trajectories over long periods of time. Stochastic systems are based on some sort of random process. They may occur when extrinsic noise is present. A process also appears to be stochastic when a large number of degrees of freedom is involved. For these systems, only statistical information can be extracted, like averages or fluctuations of physical quantities.

In this paper we consider a new example of a third, recently discovered, kind of dynamical system ({\it pseudo-deterministic}), which - in spite of being deterministic - yields only statistical relevant information. The problem with those systems is not related with the correctness and exactitude of the dynamical equations, but rather with a mathematical pathology called {\it unstable dimension variability} \cite{kurths,kostelich}.

The goodness of deterministic models is determined by some well-known paradigms: (i) the model must be based on sound theoretical framework, e.g., correctly applied physical laws; and (ii) the trajectories should reproduce correctly in some sense the actual behavior observed in Nature. This motivated the introduction of the {\it model shadowability} concept \cite{lai99}: let ${\bf A}$ and ${\bf B}$ be two closely related deterministic models of a physical system, but with some difference, which could be related to a small change in one of the system parameter values, or a slightly different external influence on each model, or a different noise level. The latter cause is restricted to arbitrarily small time dependent and bounded perturbations, which excludes Gaussian white noise, for example.

We say that model shadowability occurs if trajectories of model ${\bf A}$ stay close to some trajectories of model ${\bf B}$. This is necessary, but not sufficient, for either model to reproduce and predict correctly the time evolution of the system which the model is intended to describe. In other words, if there is no model shadowability neither ${\bf A}$ or ${\bf B}$ would generate trajectories that are physically realized, since if no trajectory of ${\bf A}$ is close to any trajectory of ${\bf B}$, it is unlikely that either model would give a trajectory that stays close to any real trajectory produced by Nature.

The difficulties that obstruct model shadowability have been divided into three classes: minor, moderate and severe \cite{lai99}. Minor modeling difficulties occur for hyperbolic chaotic systems, since they present sensitive dependence on initial conditions. In terms of our example, if ${\bf A}$ and ${\bf B}$ are hyperbolic chaotic systems, trajectories of ${\bf A}$ can always be closely followed, or shadowed, by trajectories of ${\bf B}$ for an infinite time \cite{anosov}.

A chaotic set is hyperbolic if, at each point of a trajectory on this set, the local phase space can be split into unstable and stable subspaces, the angle between them being bounded away from zero. The unstable (stable) subspace evolves into the unstable (stable) one along the trajectory. We will call {\it unstable (stable) dimension} the dimension of the unstable (stable) subspace. However, chaotic non-hyperbolic systems are much more common in physical applications - they may present non-hyperbolic (homoclinic) tangencies between the unstable and stable subspaces. For these systems we say there is a moderate modeling difficulty because trajectories of ${\bf A}$ are shadowed by trajectories of ${\bf B}$ for a long, yet finite amount of time. However, if this shadowability time is long enough, both models are still useful for describing physical phenomena \cite{shadow}.

Finally, pseudo-deterministic models present severe modeling difficulties since they are chaotic non-hyperbolic systems presenting unstable dimension variability: the unstable and stable subspaces along a chaotic invariant set have no tangencies, but the dimension of the unstable subspace varies from point to point. For this case the shadowability time is very short, and no useful information could be extracted from single trajectories over a reasonable time span, but rather statistical information based on a probability distribution \cite{kurths}.

Unstable dimension variability has been introduced by Abraham and Smale through a two-dimensional continuously differentiable map \cite{abraham}. A common feature associated with unstable dimension variability is the oscillating behavior of a finite-time Lyapunov exponent between negative and positive values. This occurs because typical trajectories present arbitrarily long (but finite-time) segments for which the orbit on the average is repelling in one dimension and other segments for which it is repelling in two dimensions. This behavior has been found in a four-dimensional invertible map describing a kicked double rotor \cite{romeiras}. The usefulness of statistical predictions has been argued through computation of the average energy and higher moments for this model \cite{kurths}.

Recently a non-invertible two-dimensional map was proposed as the simplest dynamical system exhibiting unstable dimension variability \cite{kostelich}. Moreover it has been shown \cite{lai99} that a lattice of diffusively coupled H\'enon maps present unstable dimension variability for any nonzero coupling strength.  In this paper, we propose a new example where unstable dimension variability occurs in the context of synchronization of chaotic orbits of two linear maps with nonlinear coupling. By changing variables to the synchronization manifold of the coupled system, we get a map that was first studied in the context of super-persistent chaotic transients and crises \cite{ogy83}. We have identified the mechanism which brings unstable dimension variability to the chaotic invariant set of this system, namely a saddle-repeller bifurcation which was formerly related to the boundary crisis mechanism \cite{ogy85}. It produces a structure composed of super-narrow tongues through which trajectories on a chaotic saddle may escape after very long transients before they are reinjected, since the map is taken to be modulo some number. A chaotic saddle is an invariant compact set ${\cal C}$ that is both attracting and repelling and it contains a chaotic trajectory which is dense in ${\cal C}$. This structure is similar to that observed in the context of the so-called riddling bifurcations \cite{lai96}. The fluctuating behavior of the transversal finite-time Lyapunov exponent is described for this example and statistical information about its distribution is presented.

This paper is organized as follows: in the second Section we introduce the coupled chaotic map system and analyze its synchronization manifold and the corresponding non-invertible map on a torus. The following two Sections are devoted to the description of the saddle-repeller bifurcation, the average transient lifetime  and the birth of unstable-dimension variability. Section V deals with the distribution of finite-time transversal Lyapunov exponent and the relative fraction of its positive values. The final Section contains our conclusions.

\section{Coupled Chaotic Maps}

Coupled dynamical systems are susceptible to the synchronization of their trajectories, by which they undergo closely related motions, even when they are chaotic. In the latter case, even if two identical systems are started with different initial conditions, if the coupling is strong enough their states are asymptotically equal as the time evolves \cite{pecora}. This is a quite different behavior, compared to that expected from uncoupled yet identical chaotic systems, since if they are started with approximately equal but different initial conditions, sensitive dependence will cause the two systems to have completely uncorrelated motion after some time \cite{book1}.

We consider two piecewise linear maps of the form $x_{n+1} = k x_n$ (mod $1$), where $k > 1$. For almost all trajectories of each map the (infinite time) Lyapunov exponent $\lambda = \ln k$ is positive. By {\it almost all} we mean that all orbits generated from this map are chaotic, except for a set of zero Lebesgue measure containing countably infinite periodic points \cite{book1}. We can write $k = 1 + \alpha/2$, which is greater than one provided $\alpha > 0$, and change the variable range from $[0,1)$ to $[0,\pi\sqrt{2})$, without altering the chaotic nature of the map orbits. In this case, $x$ may be regarded as an angle variable.

Let us introduce a nonlinear coupling of two such chaotic maps, which is equivalent to a nonlinear map on a torus $T^2$:

\bea
\label{mapa}
u_{n+1} = \left( 1+\frac{\alpha}{2} \right) u_n + {\cal U}(u_n,w_n) \qquad ({\rm mod } \ \pi\sqrt{2}), \\
\label{mapb}
w_{n+1} = \left( 1+\frac{\alpha}{2} \right) w_n + {\cal W}(u_n,w_n) \qquad ({\rm mod } \ \pi\sqrt{2}),
\eea

\noindent where ${\cal U}$ and ${\cal W}$ are given by

\bea
\label{U}
{\cal U}(u_n,w_n) = \left( 1-\frac{\alpha}{2} \right) w_n + \frac{1}{2\sqrt{2}} {(u_n - w_n)}^2 + \frac{\beta}{\sqrt{2}} \cos \left( \frac{u_n+w_n}{\sqrt{2}} \right), \\
\label{W}
{\cal W}(u_n,w_n) = \left( 1-\frac{\alpha}{2} \right) u_n - \frac{1}{2\sqrt{2}} {(u_n - w_n)}^2 - \frac{\beta}{\sqrt{2}} \cos \left( \frac{u_n+w_n}{\sqrt{2}} \right), 
\eea

\noindent where $\beta > 0$.

For coupled dynamical systems like this one, we can obtain the synchronization state, which is given by $u_n = w_n$. Geometrically, this state defines a synchronization manifold ${\cal S}$, which is a one-dimensional subset of the phase space $T^2 = [0,\pi\sqrt{2}) \times [0,\pi\sqrt{2})$. If we represent this torus on a square with periodic boundaries, the synchronization manifold is a straight line with unitary slope (Fig. 1).

In order to describe the dynamics in the synchronization manifold and in the direction transversal to it, we make a coordinate transformation, equivalent to a clockwise rotation of the axes through an angle of $\pi/4$:

\beq
\label{theta-z}
\theta = \frac{u+w}{\sqrt{2}}, \qquad \qquad z= \frac{u-w}{\sqrt{2}}.
\eeq

\noindent There results a two-dimensional non-invertible map on a torus 

\bea
\label{mkya}
\theta_{n+1} & = & 2\theta_n \qquad ({\rm mod } \ 2\pi), \\
\label{mkyb}
z_{n+1} & = & \alpha z_n + z_n^2 + \beta \cos\theta_n \qquad ({\rm mod } \ \pi\sqrt{2}),
\eea
\noindent where $ - \pi/\sqrt{2} < z \le + \pi/\sqrt{2} $. Here the synchronization manifold is simply the axis $z=0$. If $\beta = 0$, the map decouples into two independent maps in $\theta$ and $z$, so that an initial condition in the synchronization manifold will generate a chaotic orbit ${\{ \theta_n \}}_{n=0}^\infty$ with $z_n = 0$ for all times, that will never escape from ${\cal S}$. Hence, ${\cal S}$ is an invariant manifold only for $\beta = 0$.

An invariant manifold is typically related to the existence of some kind of symmetry in the system, so we may call $\beta$ a {\it symmetry-breaking} parameter. For nonzero $\beta$ a chaotic orbit of the system is not restricted to the synchronization manifold, and can occupy a larger portion of the phase-space along the transversal direction $z$. 

The map (\ref{mkya}-\ref{mkyb}) was introduced, in a slightly different form, \cite{ogy83} to describe a kind of crises characterized by long-lived, or super-persistent, chaotic transients. They have used the $z-$ part of the map without the modulo $\pi\sqrt{2}$ prescription, i.e., $z$ was allowed to have any real value in $(-\infty,+\infty)$. For some parameter values, as $\alpha = 0.7$ and $\beta = 0.02$, the map (\ref{mkya}-\ref{mkyb}) shows a chaotic attractor near the $z=0$ line. It consists of a series of curved strips with self-similar structure, as H\'enon-like bands (Fig. 2) \cite{pim}. A crude argument that can be used to justify the chaotic nature of this attractor consists on linearizing the map (\ref{mkyb}) about $z=0$ using $z = \beta \zeta$, where $\zeta$ is a small quantity. We obtain the map $\zeta_{n+1} = \alpha \zeta_n + \cos\theta_n$, that would give a chaotic attractor for $|\alpha| < 1$, as proved by Kaplan and Yorke \cite{kaplan}.

If the modulo $\pi\sqrt{2}$ prescription were absent from $z$-map, the system would present, in addition, a non-chaotic attractor at infinity $(z = +\infty)$, and the basin of attraction of the two attractors would present a fractal boundary \cite{ogy83}. In our case, however, there is no other attractor except the one near $z = 0$. Besides this chaotic attractor, there is another chaotic set which is a chaotic saddle (a non-attracting chaotic invariant set). It is located where the fractal basin boundary would be if the modulo requirement were disregarded. Trajectories which would accelerate to large-$z$ values, are reinjected into the negative-$z$ region due to the modulo prescription.

\section{Saddle-Repeller Bifurcation}

The mechanism whereby the chaotic attractor of the map (\ref{mkya}-\ref{mkyb}) loses hyperbolicity through unstable dimension variability is basically an unstable-unstable pair bifurcation. As a result, the chaotic attractor may collide with the chaotic saddle and disappears into a larger chaotic saddle from which trajectories may escape, through a complex structure of super-narrow tongues. It is also called saddle-repeller bifurcation \cite{ogy85}, and has been found as the cause of other strange behavior in chaotic systems, like riddling of basins of attraction \cite{lai96} and boundary crises \cite{ogy83}.

A linear stability analysis can show the basic features of this transition. The period-1 fixed points of the map (\ref{mkya}-\ref{mkyb}) are

\beq
\label{fixed}
{\bar\theta} = 0, \qquad \qquad {\bar z} = \frac{1}{2} \left( 1-\alpha \pm \sqrt{
{(1-\alpha)}^2 - 4\beta} \right).
\eeq

\noindent Defining

\beq
\label{star}
z_* = \frac{1-\alpha}{2}, \qquad \qquad \beta_* = z_*^2,
\eeq

\noindent these fixed points are written as 

\beq
\label{zbc}
(\theta = 0, z_b = z_* + \sqrt{\beta_* - \beta}), \qquad {\rm and} \qquad
(\theta = 0, z_c = z_* - \sqrt{\beta_* - \beta}).
\eeq

\noindent Let us fix our attention to the case depicted in Fig. 2, i.e., $\alpha = 0.7$ and $\beta = 0.02 < \beta_* = 0.0225$. It turns out that $z_c$ lies in the upper point (i.e., the point with the highest value of $z$) of the chaotic attractor, whereas $z_b$ is in the lowest point of the chaotic saddle.

The Jacobian matrix of the map is

\beq
\label{J}
\left({
\begin{array}{c c}
2 & 0 \\
- \beta \sin\theta_n & \alpha + 2 z_n 
\end{array}
}\right),
\eeq

\noindent whose eigenvalues are $\xi_1 = 2$ and $\xi_2 = \alpha+2 z_n$. So, the $\theta$-direction is always unstable, as it should be due to the existence of the chaotic attractor. The eigenvalue related to the transversal direction, evaluated at the fixed points (\ref{zbc}), gives

\beq
\label{autovalor}
\xi_2(z_{b,c}) = 1 \pm 2 \sqrt{\beta_* - \beta},
\eeq

\noindent so that, for $\beta < \beta_*$, $z_b$ is a repeller, since it has an unstable dimension of dimension 2, and no stable subspace at all. $z_c$ is a saddle point, since both stable and unstable subspaces have dimensions equal to one (Fig. 3(a)). 

If $\beta = \beta_*$ the fixed points $z_c$ and $z_b$ coalesce at $z = z_*$ (Fig. 3(b)). Since the eigenvalue is equal to one at this point, linear stability analysis fails to determine its stability, and we have a saddle-repeller bifurcation with eigenvalue $+1$ at $\beta = \beta_*$. For $\beta > \beta_*$ the fixed points no longer exist, and we have a different dynamical behavior (Fig. 3(c)).

When the fixed points coalesce at $\beta = \beta_*$, the chaotic attractor collides with the chaotic saddle, and it becomes a {\it chaotic transient} through an interior crisis (since the attractor has collided with an unstable periodic orbit) \cite{ogy85}. Trajectories arising from initial conditions belonging to the former basin of the chaotic attractor will typically approach its remnant, that is now a part of a larger chaotic saddle and will bounce around it in an irregular fashion. However, after some (typically very long) time this trajectory will stay near the region where the fixed points have coalesced, and will leave the chaotic saddle, being eventually reinjected to the vicinity of the saddle due to the modulo prescription.

At the location where the fixed points coalesced $(z = z_*)$, a tongue opens-up, allowing trajectories near the chaotic saddle to escape for $\beta > \beta_*$. Simultaneously, each preimage of $z_*$ also develops a tongue. Since these preimages are dense in the chaotic saddle, an infinite number of these tongues open-up simultaneously when $\beta = \beta_*$. Actually, these tongues will develop at those points where $\theta = 2\pi m / 2^k$, with $m$ and $k$ positive integers. For our map, however, these tongues are very narrow, since their width decrease geometrically, and are extremely difficult to find numerically, with exception of the main tongue opened-up at $z_*$. 

\section{Behavior of Chaotic Transients}

To understand how chaotic transients are formed after the saddle-repeller bifurcation, let us consider the z-part of the map (\ref{mkyb}) at $\theta = 0$ (from where the main tongue grows up)

\beq
\label{zmap}
z_{n+1} = \alpha z_n + z_n^2 + \beta,
\eeq

\noindent whose fixed points $z_b$ and $z_c$ are the intersections between the parabolic function and the first bisector $z_{n+1}=z_n$ (Fig. 4(a)). As $\beta$ approaches its critical value $\beta_*$, these points approach each other and eventually coalesce when $\beta = \beta_*$ at $z=z_*$, where the map function is tangent to the $45^o$-line (Fig. 4(b)). For $\beta > \beta_*$ the parabola has moved upwards and does not intercept the bisector, leaving no fixed points. However, provided $\beta$ is not much far from $\beta_*$, a narrow channel forms between the parabola and the $45^o$-line, through which passes the trajectory resulting from the map iterations, staying there a very long time, and eventually escaping to large $z$-values (Fig. 4(c)). Due to the modulo requirement, the map iterates are reinjected and enter again in the channel. This is the basic mechanism of Type-I Intermittency \cite{pomeau}, but since the $\theta$-direction is unstable, the slow motion through the channel does not imply a laminar behavior. Rather, it is related to a chaotic transient, that is characterized by irregular wandering of the trajectory over a limited $\theta$-range.

The chaotic transient decays when the trajectory enters a tongue, escaping toward larger $z$-values, before it is reinjected. The main tongue is formed about the $\theta = 0$ line, so let us focus our attention on this spot. For $\beta$ values greater than $\beta_*$, a tongue intercepts the synchronization manifold $z = 0$ in an aperture of width $\ell$ near $\theta = 0$, through which the trajectories can escape. Once having entered the aperture, the trajectory stays in it a number $T$ of iterates before leaving the region $y \le y_c$. Although $y_c$ must be less than $\pi/\sqrt{2}$, its exact value does not affect our results in a significant way. From now on, we set $y_c = 2.0$. The trajectory, however, does not immediately stays close to $\theta = 0$ since the $\theta$-direction is unstable (its eigenvalue is $\xi_2 = 2 > 1$, pulling back the trajectory into the vicinity of the chaotic saddle).

Let $\ell$ be the distance between the orbit and $\theta = 0$ at a given time. After $T$ iterates this distance increases to ${(\xi_2)}^T \ell$. However, since the $\theta$-excursion is bounded, we expect that ${(\xi_2)}^T \ell < \kappa_1$, where $\kappa_1$ is a ${\cal O}(1)$ constant. Now we consider what happens with a large number of trajectories arising from randomly chosen initial conditions. Due to the ergodicity of the $\theta$-motion, we expect that almost any initial condition $\theta_0$ will generate an orbit which has a uniform distribution over $[0,2\pi)$ for large times. Hence, the probability of $\theta_{n+1}$ falling into the aperture $[-\ell/2,+\ell/2]$, if $\theta_n$ is not in this interval, is equal to the interval width $\ell$. So, we can estimate the average lifetime $<\tau>$ of the chaotic transient as the inverse of this probability. Taking our previous estimate as an upper bound for the distance along $\theta$-direction at time $T$, we have \cite{ogy85}

\beq
\label{tau}
<\tau> = \kappa {(\xi_2)}^T = \kappa  2^T,
\eeq

\noindent where $\kappa = 1/\kappa_1$.

In order to compute $T$, we assume that, for $\beta$ slightly greater than $\beta_*$ the aperture width $\ell$ is very small, so that we may approximate $\theta$ by $0$ and use Eq. (\ref{zmap}) again. The difference $z_{n+1} - z_n$ has a local minimum at $z_*$, so we describe the dynamics within the narrow channel by using $\delta_n = z_n - z_*$. This difference evolves with time according to the map

\beq
\label{delta}
\delta_{n+1} = \delta_n + \delta_n^2 + (\beta - \beta_*),
\eeq

\noindent and since $\delta_{n+1}$ is very close to $\delta_n$ we can approximate the difference $\delta_{n+1}-\delta_n$ as a differential and write a differential equation for $\delta$, now a continuous function of time $\tau$:

\beq
\label{delta2}
\frac{d\delta}{d\tau} = \delta^2 + (\beta - \beta_*),
\eeq

\noindent which can be integrated from $-\pi/\sqrt{2}$ to $+\pi/\sqrt{2}$, to give an estimate for the time $T$ it takes for the trajectory to escape once it has fallen in the aperture: 

\beq
\label{T}
T = \frac{2}{\sqrt{\beta - \beta_*}} \arctan \left( \frac{\pi/\sqrt{2}}{\sqrt{\beta - \beta_*}} \right).
\eeq

\noindent Inserting Eq. (\ref{T}) into Eq. (\ref{tau}), and taking logarithms we have

\beq
\label{theory}
\log_{10} <\tau> = \log_{10} \kappa + 2 \log_{10} 2 {(\beta - \beta_*)}^{-1/2} 
\arctan \left( \frac{\pi}{\sqrt{2}}{(\beta - \beta_*)}^{-1/2} \right).
\eeq

To check the validity of the hypotheses made in the above derivation, we have made a numerical experiment, choosing $N_{\theta_0}$ initial conditions randomly distributed over $[0,2\pi)$, and computing the transient exit time once a given trajectory crosses the line $y = y_c = 2.0$. The average transient lifetime $<\tau>$ was then computed for many values of the difference ${(\beta - \beta_*)}^{-1/2}$, the results being depicted in Fig. 5, where the parameter $\kappa$ in the theoretical prediction above is the fitting parameter.

The fitting is asymptotic though, since numerical results are best fitted by Eq. (\ref{theory}) when we approach $\beta_*$, i.e., in the neighborhood of the saddle-repeller bifurcation. The lifetime of the transients can be very large, for instance up to $10^8$ transients, even though we are as far from $\beta_*$ as $44\%$. This occurs because the aperture width $\ell$ is so narrow that it becomes extremely difficult for an orbit to escape. Therefore, a better agreement between the theoretical prediction (\ref{theory}) and the numerical result only occurs for $\beta$ closer to $\beta_*$. Far from $\beta_*$, the theoretical estimate we made ceases to be valid, but remains useful as a lower bound for the exit time of transients. 

\section{Finite-Time Lyapunov Exponents}

We have seen that, at the saddle-repeller bifurcation $\beta = \beta_*$, an infinite number of points, dense on the chaotic attractor, become repellers (unstable dimension two), and an infinite number of super-narrow tongues crop up as result of the collision between the attractor and the chaotic saddle. However, there remains a dense set of saddle points (unstable dimension one) in the invariant set, and since these two different sets are densely intertwined, unstable dimension variability does occur in the chaotic invariant set for a large range of parameter values of the map.

Yet, another signature of unstable dimension variability, as stated in the Introduction, is the fluctuating behavior of the finite-time Lyapunov exponents between negative and positive values. Consider a d-dimensional map ${\bf x} \rightarrow {\bf f}({\bf x})$, where ${\bf x}$ is a d-dimensional vector and ${\bf f}$ is a d-dimensional vector field. Let $n$ be a positive integer and let ${\bf Df}^n({\bf x}_0)$ denote the Jacobian matrix of ${\bf f}^n$ (the n-times iterated map function) evaluated at the point ${\bf x}_0$. The eigenvalues of the Jacobian matrix ${\bf Df}^n({\bf x}_0)$ are

\beq
\sigma_1({\bf x_0},n) \ge \sigma_2({\bf x_0},n) \ge \ldots  \ge \sigma_d({\bf x_0},n) \ge 0.
\eeq

\noindent We define the k-th time-n Lyapunov exponent associated with the initial condition ${\bf x}_0$ as \cite{kostelich}

\beq
\lambda_k({\bf x}_0,n) = \frac{1}{n} \ln || {\bf Df}^n({\bf x}_0) {\bf u}_k||,
\eeq

\noindent where ${\bf u}_k$ is the eigenvector corresponding to the eigenvalue $\sigma_k$. Note that the usual infinite-time Lyapunov exponent

\beq
\lambda_k = \lim_{n\rightarrow\infty} \lambda_k ({\bf x}_0,n)
\eeq

\noindent has the same value for almost every initial point ${\bf x}_0$ with respect to the Lebesgue measure in the basin of attraction.

For our map on a 2-torus, ${\bf x}_n = {(\theta_n,z_n)}^T$, the Jacobian matrix of the n-times iterated map has eigenvalues $\sigma_1 = {(\xi_1)}^n = 2^n$, and 

\beq
\label{eigenvalue}
\sigma_2 = \prod_{i=1}^n \alpha + 2 z_i,
\eeq

\noindent so that the second (transverse) finite-time Lyapunov exponent is given by 

\beq
\label{lambda2}
\lambda_2({\bf x}_0,n) = \frac{1}{n} \sum_{i=1}^n \ln \vert \alpha+2z_i \vert,
\eeq

\noindent the first exponent being simply $\ln 2$.

The possibility of fluctuations between positive and negative values for this exponent makes useful to define a distribution function for it. Let $P(\lambda_2({\bf x}_0,n),n)$ denote the probability density function of the second time-n Lyapunov exponent, when ${\bf x}_0$ is chosen at random according to the Lebesgue measure. In other words, $P(\lambda_2({\bf x}_0,n),n) d\lambda_2$ is the probability that the exponent value lies between $\lambda_2$ and $\lambda_2 + d\lambda_2$. If $F(\lambda_2)$ is any function of the time-n Lyapunov exponent, its average over the invariant measure of the attractor is given by

\beq
\label{average}
\left\langle F(\lambda_2({\bf x}_0,n) \right\rangle = \int F(\lambda_2({\bf x}_0,n)) P(\lambda_2({\bf x}_0,n),n) d\lambda_2
\eeq

To obtain the distribution $P(\lambda_2)$ numerically, we picked many randomly chosen initial conditions uniformly distributed in $[0,2\pi)$, and iterate each initial condition ${\bf x}_0$ a large number of times. Every $n = 10$ steps we compute the time-10 exponent according to Eq. (\ref{lambda2}). Actually, we use the recurrency of dynamics and follow a single trajectory a large number of steps, say 2 million iterates. The time $n=10$ exponents are then computed from $2 \times 10^5$ consecutive and non-overlapping length-$10$ sections of the trajectory. From these exponents we compute a frequency histogram with convenient normalization for that

\beq
\label{norm}
\int_{-\infty}^{+\infty} P(\lambda_2({\bf x}_0,n),n) d\lambda_2 = 1.
\eeq

In Figure 6(a), we show a distribution for $\alpha = 0.7$ and $\beta = 0.04$, which is $\approx 78 \%$ away from the critical value $\beta_* = 0.0225$. In this case we can observe a distribution which looks like a Gaussian, but with asymmetric tails. The negative tail is sharply cut off, whereas the positive tail decreases smoothly. Only $0.23 \%$ of the second finite-time exponents are positive, indicating that almost all trajectory sections are transversally contracting. This is consistent with the trajectory behavior in the narrow channel that occurs near $\beta_*$, but the noteworthy feature here is the relatively small number of positive exponents. Figure 6(b) depicts the same situation, but for $\beta = 0.07$, which is about three times the previous deviation away from the critical value. This time the Gaussian character of the distribution is apparent only for negative exponents, while there is a long, non-Gaussian flat tail of positive exponents ($9 \%$ of their total number). The maximum of the distribution, however, appears not to have moved either toward less negative or positive  values of $\lambda_2(10)$, having approximately the same value of $-0.35$ in both cases. The fate of the probability distribution as we increase further the symmetry-breaking parameter $\beta$ is illustrated in Fig. 6(c), where we used $\beta = 0.15$. The same general characteristics of the previous figure are still here, even though we are now very far (almost six times) from the critical value. The negative peak still exists at approximately the same location, but the distribution has broadened in that place. The novel aspect is the emergence of a second peak in the positive tail. We note that the fraction of positive exponents has increased to $37\%$.

The fraction of positive time-n exponents

\beq
\label{f}
f(n) = \int_0^\infty P(\lambda_2({\bf x}_0,n), n) d\lambda_2
\eeq

\noindent has been computed for various values of $\beta$, the results being depicted in Fig. 7, showing a monotonic increase of this fraction, indicating that for $\beta \approx 0.17$ about $40 \%$ of the exponents are positive. However, this number increases in the map (\ref{mkya}-\ref{mkyb}) due to the emergence of the second peak in the positive tail of the distribution. The shape of the curve in Fig. 7 strongly suggests a kind of integrated probability distribution. We have thus computed the cumulative histogram

\beq
\label{cumul}
Q(\lambda_2, n) = \int_{-\infty}^{\lambda_2} P(\lambda_2^{'},n) d\lambda_2^{'} = 1 - \int_{\lambda_2}^\infty P(\lambda_2^{'},n) d\lambda_2^{'},
\eeq

\noindent where we have used Eq. (\ref{norm}). In this case $Q(\lambda_2\rightarrow -\infty) = 0$ and $Q(\lambda_2\rightarrow +\infty) = 1$. In Fig. 8 we show a cumulative histogram related to the distribution depicted in Fig. 6(c), i.e., for $\beta = 0.15$. In fact, if we compare it with the previous figure, qualitatively they are very similar, since for positive exponents it deviates from an integral of a Gaussian shaped function. Using Eq. (\ref{cumul}), it is easy to show that $f(n) = 1 - Q(0,n)$.

\section{Conclusions}

In summary, this work proposes a theoretical mechanism for the existence of unstable dimension variability in synchronized coupled chaotic systems, which is confirmed by analysis and numerical computation of finite-time Lyapunov exponents. 

For invariant sets of a dynamical system, unstable dimension variability can be a strong obstacle to mathematical modeling of physical phenomena, since there is little probability that a real chaotic trajectory is shadowed for moderately long times by a trajectory from a model. This puts some serious doubts on the deterministic model itself, not because it is intrinsically bad, but rather because it presents a pathology that prevents adequate model shadowability. The consequence is that, although the model is deterministic, we expect to make only relevant statistical predictions, like averages or fluctuations, based on it.

\section*{Acknowledgments}

This work was made possible by partial financial support from National Science Foundation (Division of International Programs) and CNPq (Conselho Nacional de Desenvolvimento Cient\'{\i}fico e Tecnol\'ogico) - Brazil, through a joint research collaboration project. It was also supported in part by ONR (Physics).

\newpage

\begin{center}
{\bf FIGURE CAPTIONS}
\end{center}

FIGURE 1: Synchronization manifold and its transversal direction for a 2-torus.
 
FIGURE 2: Phase portrait of the map (\ref{mkya}-\ref{mkyb}) for $\alpha = 0.7$ and $\beta = 0.02 < \beta_* = 0.0225$. The dark region contains points which are driven to higher $z-$ values but are eventually reinjected to negative $z$-values due to the modulo requirement. The chaotic saddle is the boundary between the dark and light grey region. The chaotic attractor is embedded in the light grey region.

FIGURE 3: Fixed points for the map (\ref{mkya}-\ref{mkyb}) and their stability for (a) $\beta < \beta_*$, (b)  $\beta = \beta_*$, and (c) $\beta > \beta_*$.

FIGURE 4: Return map for Eq. (\ref{mkyb}) with $\theta = 0$ and $\alpha = 0.7$ for (a) $\beta = 0.01 < \beta_* = 0.0225$, (b)  $\beta = \beta_* = 0.0225$, and (c) $\beta = 0.04 > \beta_*$.

FIGURE 5: Base-10 logarithm of the average exit time of transients $<\tau>$ of the map  (\ref{mkya}-\ref{mkyb}) with $\alpha = 0.7$ and $\beta_* = 0.0225$ as a function of ${(\beta-\beta_*)}^{-1/2}$. Circles: numerical experiment, full line: theoretical prediction given by Eq. (\ref{theory}) with a fitting parameter $\kappa = 0.0045$.

FIGURE 6: Probability distribution for the transverse time-$10$ Lyapunov exponents for $\alpha = 0.7$ and (a) $\beta = 0.04$; (b) $\beta = 0.07$; and (c) $\beta = 0.15$.

FIGURE 7: Fraction of positive transverse time-$10$ Lyapunov exponents $f(n = 10)$ as a function of $\beta$, for $\alpha = 0.7$ and $\beta_* = 0.0225$.

FIGURE 8: Cumulative histogram for the transverse time-$10$ Lyapunov exponents for $\alpha = 0.7$ and $\beta = 0.15$.

\newpage

\begin{figure}[htbp]
\begin{center}
\leavevmode
\hbox{
\epsfxsize=12.0cm
\epsffile{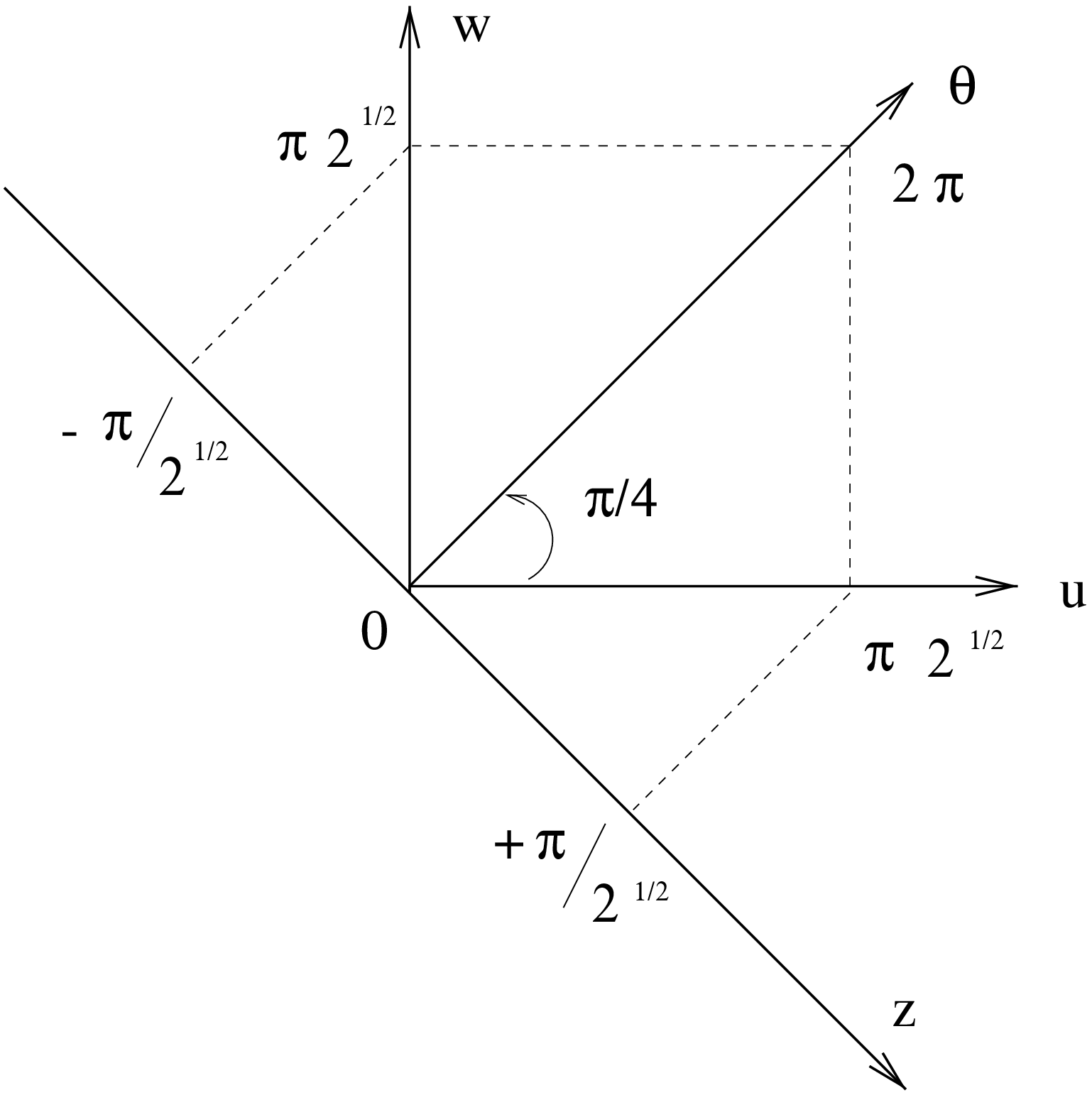}}
\caption{}
\end{center}
\end{figure}

\newpage
\begin{figure}[htbp]
\begin{center}
\leavevmode
\hbox{
\epsfxsize=12.0cm
\epsffile{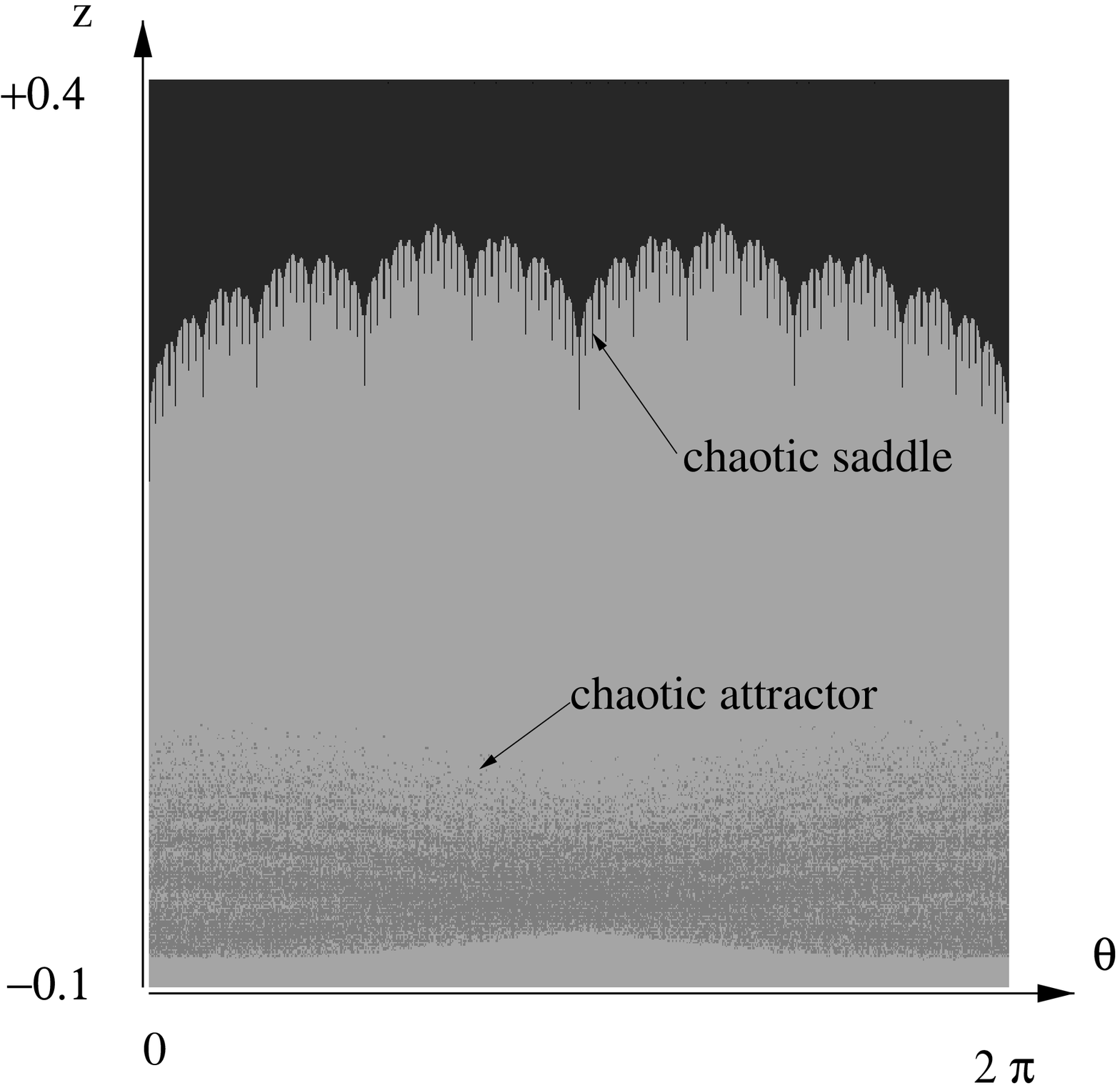}}
\caption{}
\end{center}
\end{figure}

\newpage
\begin{figure}[htbp]
\begin{center}
\leavevmode
\hbox{
\epsfxsize=13.0cm
\epsffile{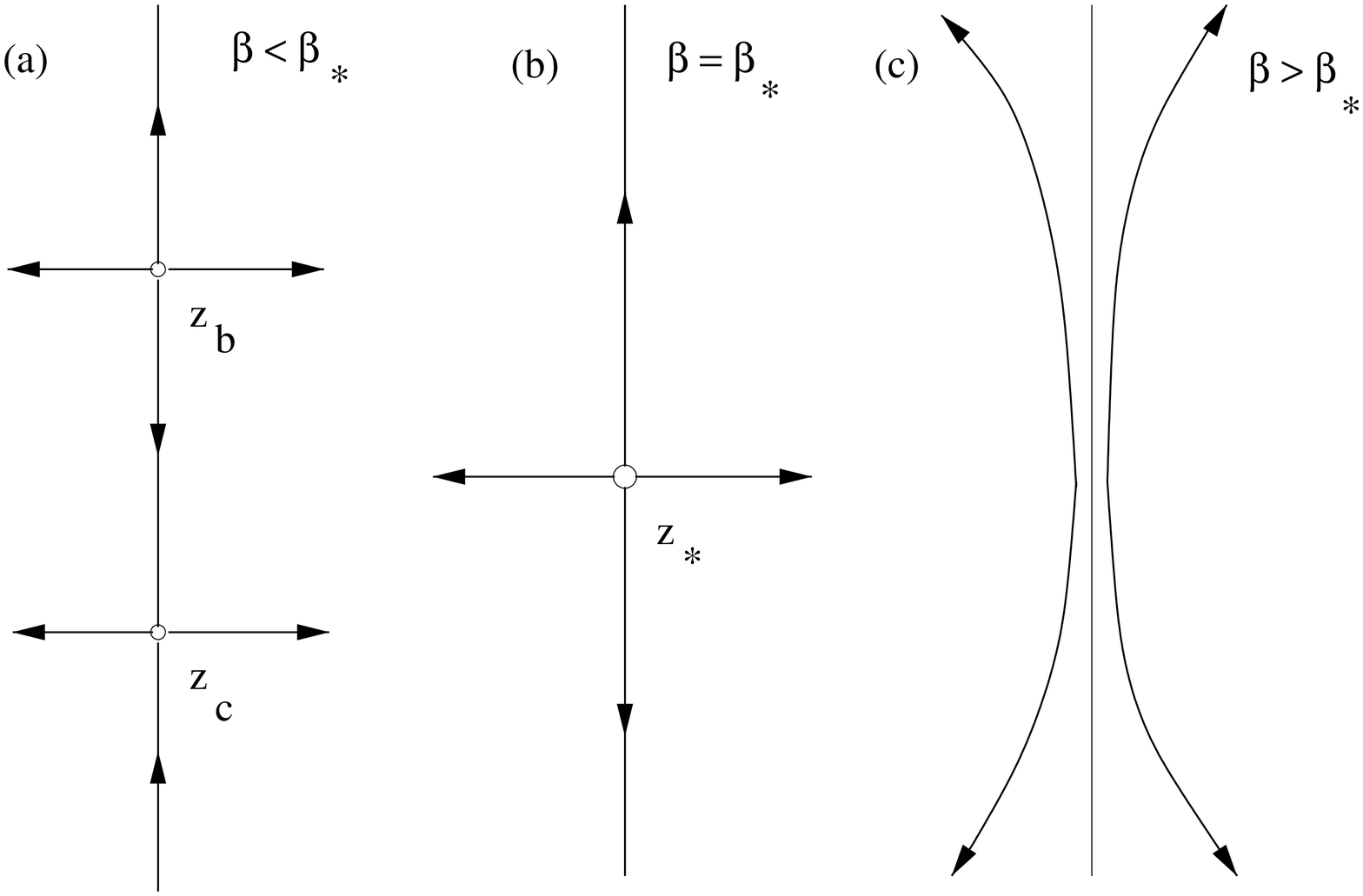}}
\caption{}
\end{center}
\end{figure}

\newpage
\begin{figure}[htbp]
\begin{center}
\leavevmode
\hbox{
\epsfxsize=12.0cm
\epsffile{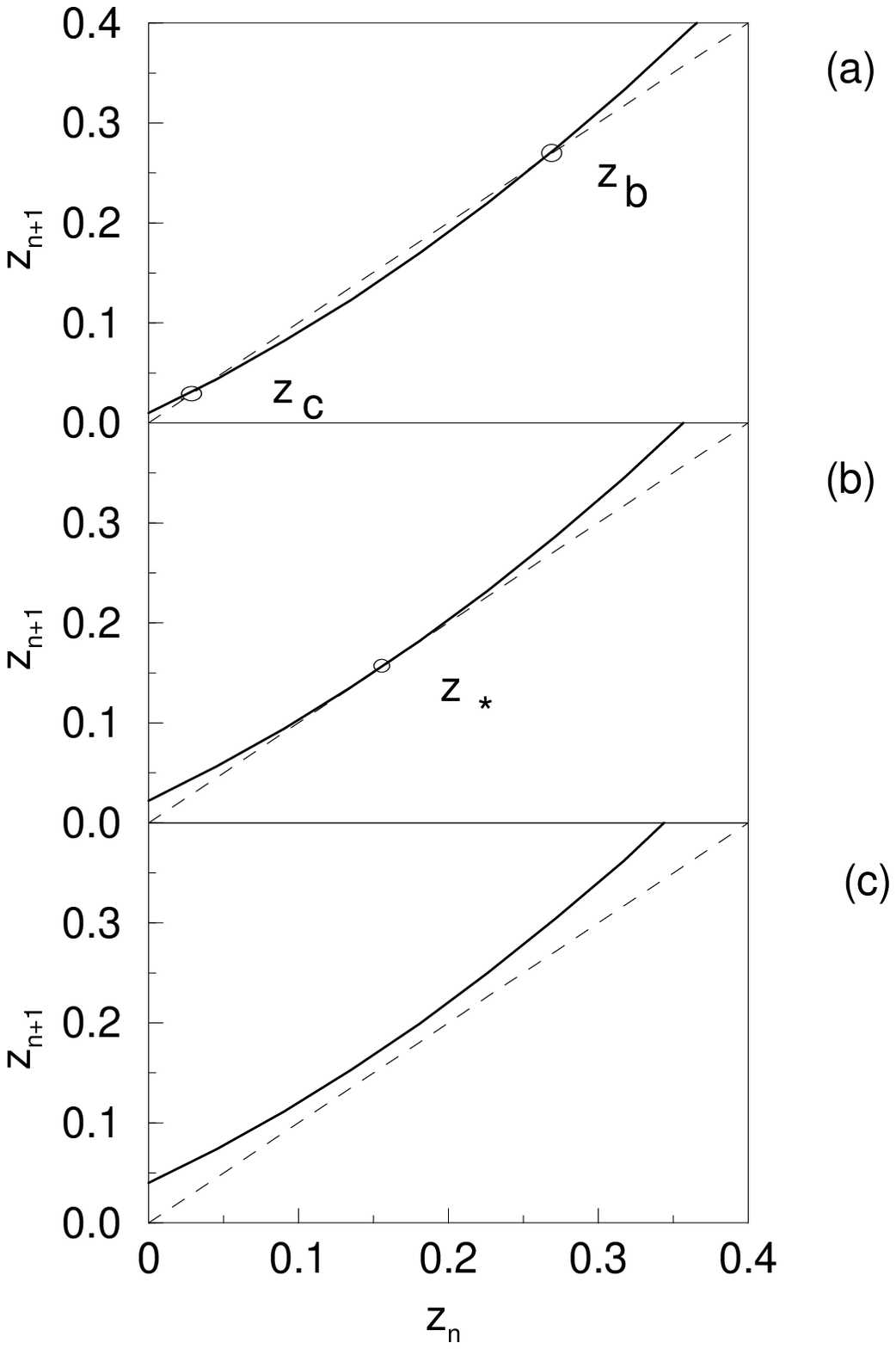}}
\caption{}
\end{center}
\end{figure}

\newpage
\begin{figure}[htbp]
\begin{center}
\leavevmode
\hbox{
\epsfxsize=13.0cm
\epsffile{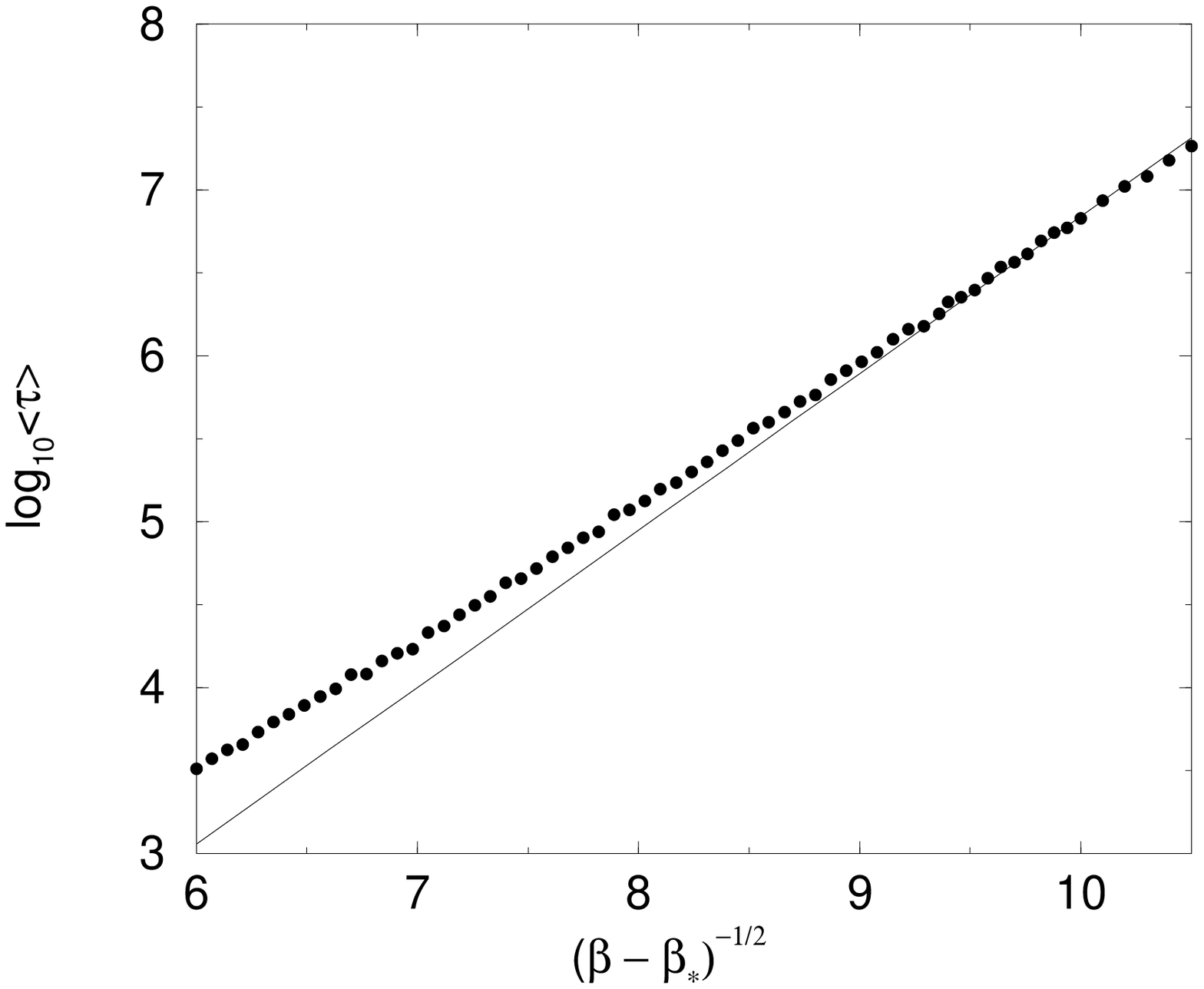}}
\caption{}
\end{center}
\end{figure}

\newpage
\begin{figure}[htbp]
\begin{center}
\leavevmode
\hbox{
\epsfysize=15.0cm
\epsffile{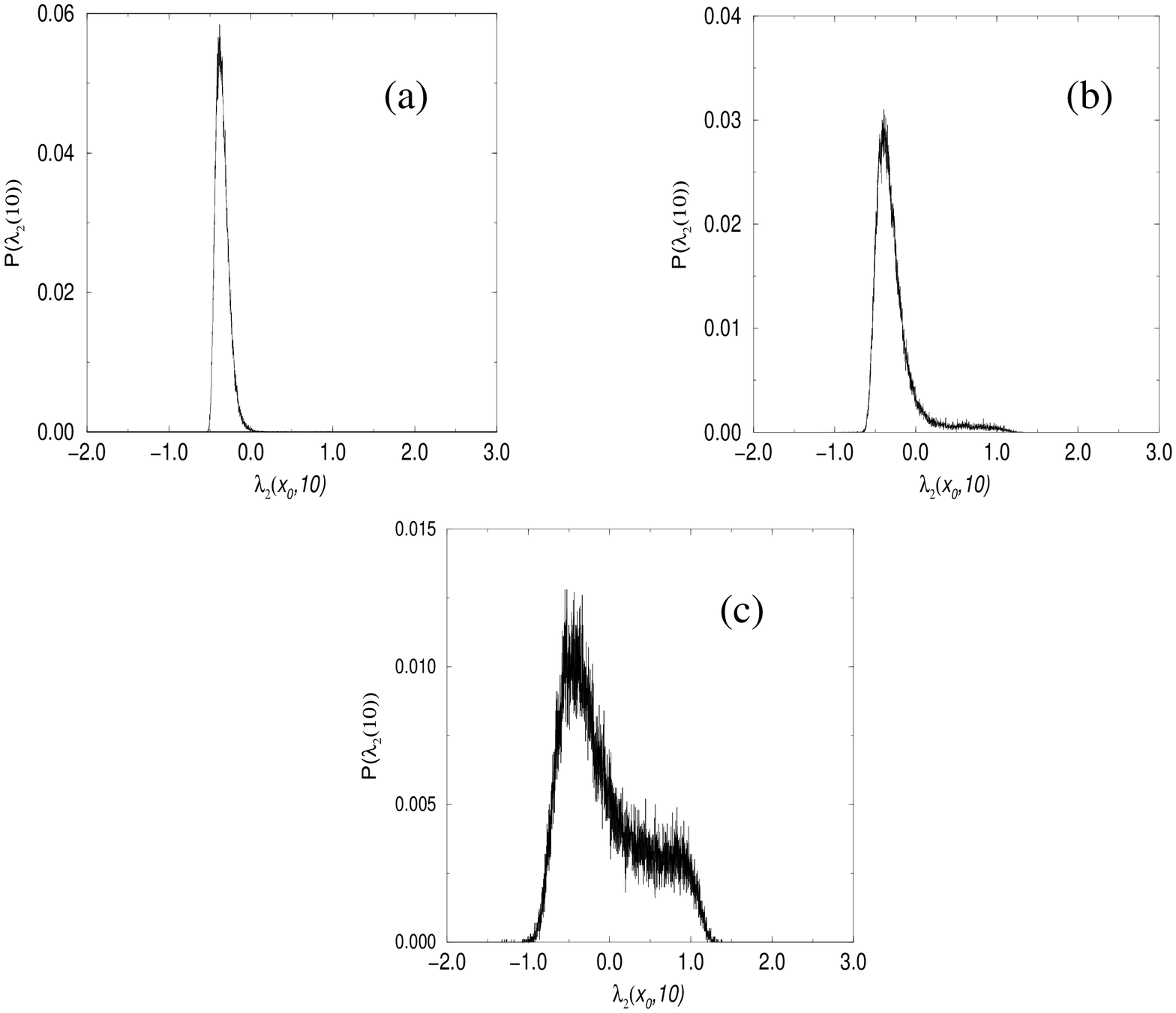}}
\caption{}
\end{center}
\end{figure}

\newpage
\begin{figure}[htbp]
\begin{center}
\leavevmode
\hbox{
\epsfxsize=13.0cm
\epsffile{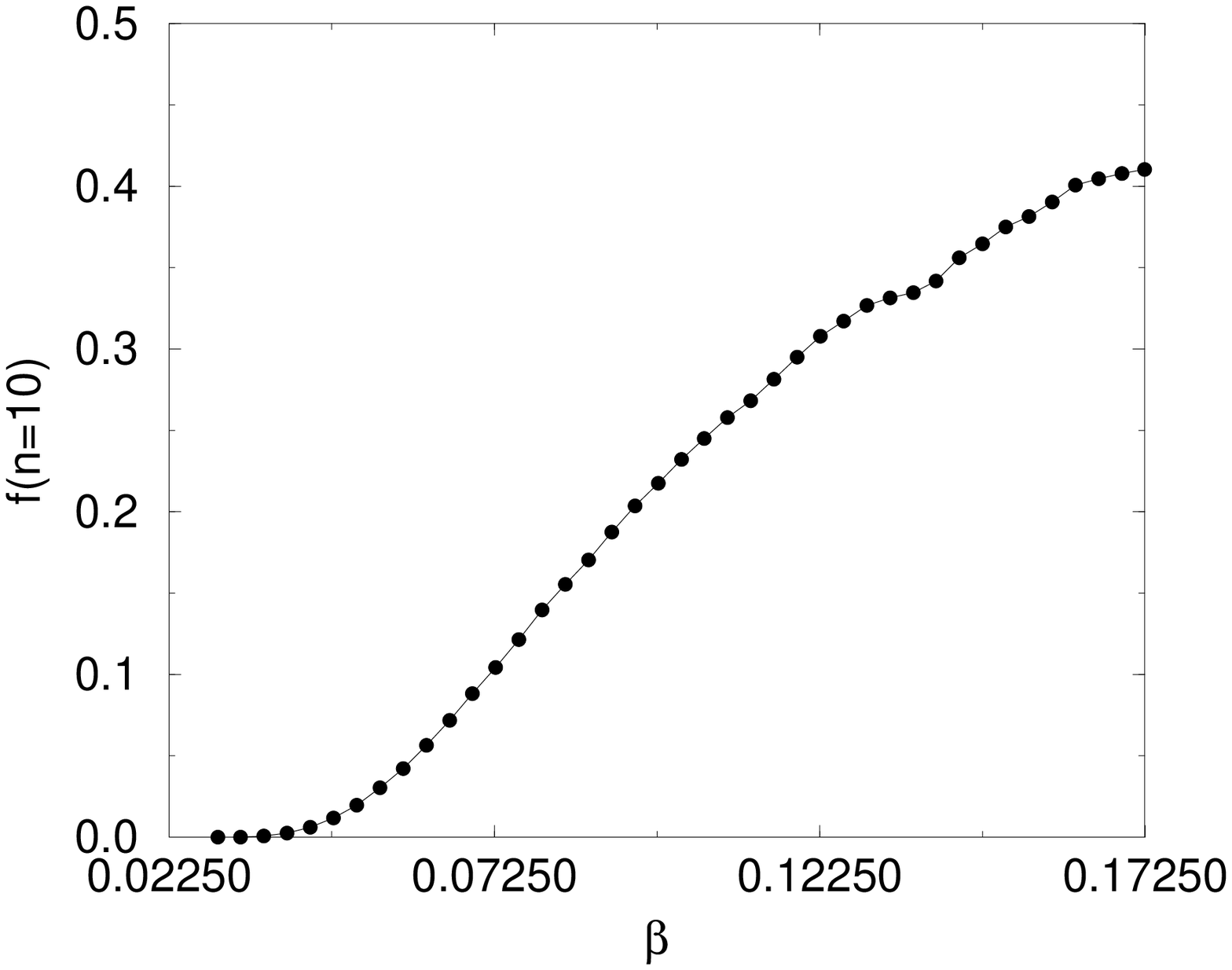}}
\caption{}
\end{center}
\end{figure}

\newpage
\begin{figure}[htbp]
\begin{center}
\leavevmode
\hbox{
\epsfxsize=13.0cm
\epsffile{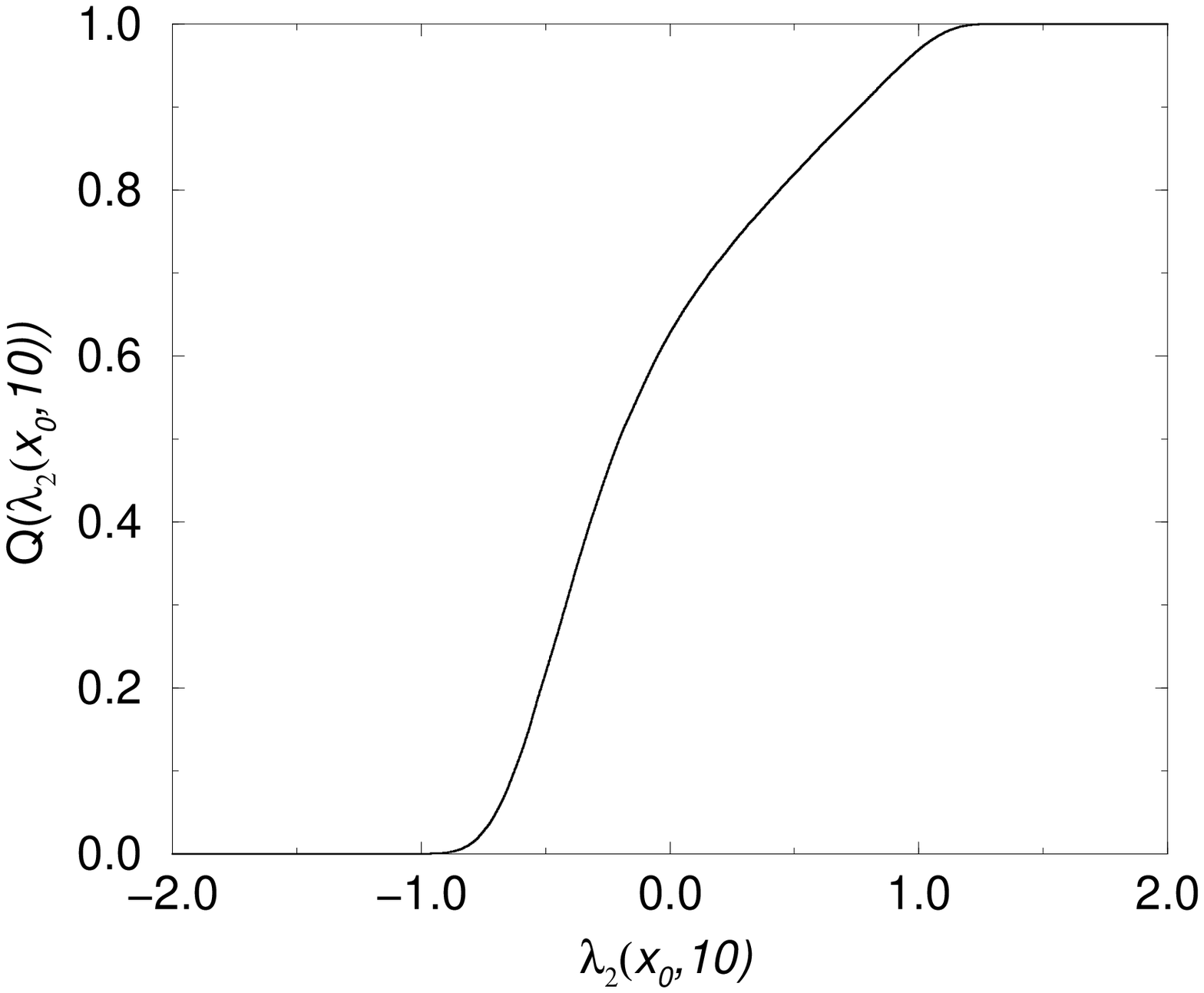}}
\caption{}
\end{center}
\end{figure}

\end{document}